\documentclass[9pt,twocolumn,twoside]{osajnl}
\journal{ol}
\usepackage[T1]{fontenc}
\usepackage{float}
\usepackage{graphicx}% http://ctan.org/pkg/graphicx
\setboolean{shortarticle}{true} 

\ifthenelse{\boolean{shortarticle}}{\colorlet{color2}{color2b}}{\colorlet{color2}{color2}} 

\title{Wavefront-sensing with a thin diffuser}%, exploiting its memory effect}

\author[1]{Pascal Berto}
\author[2]{Herv\'{e} Rigneault} 
\author[1,*]{Marc Guillon}

\affil[1]{Sorbonne Paris Cité, Université Paris Descartes, Neurophotonics Laboratory, CNRS-UMR 8250, 45 rue des Saints-Pères, F-75006 Paris, France}
\affil[2]{Aix Marseille Univ, CNRS, Centrale Marseille, Institut Fresnel, Marseille, France}

\affil[*]{Corresponding author: marc.guillon@parisdescartes.fr}

\dates{Compiled \today}

\ociscodes{(120.5050) Phase measurement; (110.0113) Imaging through turbid media; (180.3170) Interference microscopy.}
\doi{\url{http://dx.doi.org/10.1364/ao.XX.XXXXXX}}

\begin{abstract}
We propose and implement a broadband, compact, and low-cost wavefront sensing scheme by simply placing a thin diffuser in the close vicinity of a camera. The local wavefront gradient is determined from the local translation of the speckle pattern. The translation vector map is computed thanks to a fast diffeomorphic image registration algorithm and integrated to reconstruct the wavefront profile. The simple translation of speckle grains under local wavefront tip/tilt is ensured by the so-called ``memory effect'' of the diffuser. Quantitative wavefront measurements are experimentally demonstrated both for the few first Zernike polynomials and for phase-imaging applications requiring high resolution. We finally provided a  theoretical description of the resolution limit that is supported experimentally.
\end{abstract}

\setboolean{displaycopyright}{true}

\begin{document}

\maketitle
\thispagestyle{fancy}
\ifthenelse{\boolean{shortarticle}}{\ifthenelse{\boolean{singlecolumn}}{\abscontentformatted}{\abscontent}}{}
%
%But WFS still expensive cause to alignment and phase mask engineering (Hartmann Mask\cite{Vdovin_OE_16} or microlenses)
%%Comparison with classical Phase sensitive: DIC, Phase retrieval... - not quantitative or not one shot...
%Diffuser based camera (Laura Waller): https://www.osapublishing.org/abstract.cfm?uri=COSI-2017-CM2B.2
%Off-axis digital holographic camera for quantitative phase microscopy:\cite{Moser_BOE_14}
%Compact lensless phase imager \cite{Moser_OE_17}

The phase map of a light beam can be typically determined by digital holography~\cite{Cuche1999}, in which case the phase is defined relatively to a reference beam. To minimize the phase noise associated with accidental shifts and instabilities between the reference and the signal beam, inline holography~\cite{Gabor_PRSLA_49} and common path interferometers~\cite{Anderson_AO_95,Gluckstad_OC_96,Marte_OL_05,Guillon_JOSA_14} were developed. However, the reference wave is then extracted from the signal wave, making its amplitude object-dependent. 
%Wavefront sensing necessary broadbad??? Not sure at all.
%Starting by phase measure by interferometry 
Alternatively, to avoid the need for a reference beam, the local light wavevector can be directly mapped over a two dimensional space allowing to determine the beam wavefront. With such wavefront sensors (WFS), the phase is recovered by integrating the wavevector map, up to an integration constant accounting for the absence of phase reference. One of the main advantages of direct wavefront measurement %as compared to holographic techniques 
is that it is easily compatible with broadband light sources. Potentially, the local intensity can also be measured, so-giving access to the complex amplitude of the field. First applied for aberration compensation by adaptive optics in astronomy~\cite{Roddier_PASP_91}, WFS has also been of primary interest for optical metrology, laser beam characterization, ophtalmology, and more recently for imaging through complex media~\cite{Denk_PNAS_06}. High resolution WFS now makes  quantitative phase imaging possible for cell microscopy~\cite{Bon2009}, microthermometry~\cite{Baffou2012}, or vibrational and absorption microspectroscopy~\cite{Berto2012b,Berto2012a, Berto2013}.

\begin{figure}[htbp]
\centering
\includegraphics[width=\linewidth]{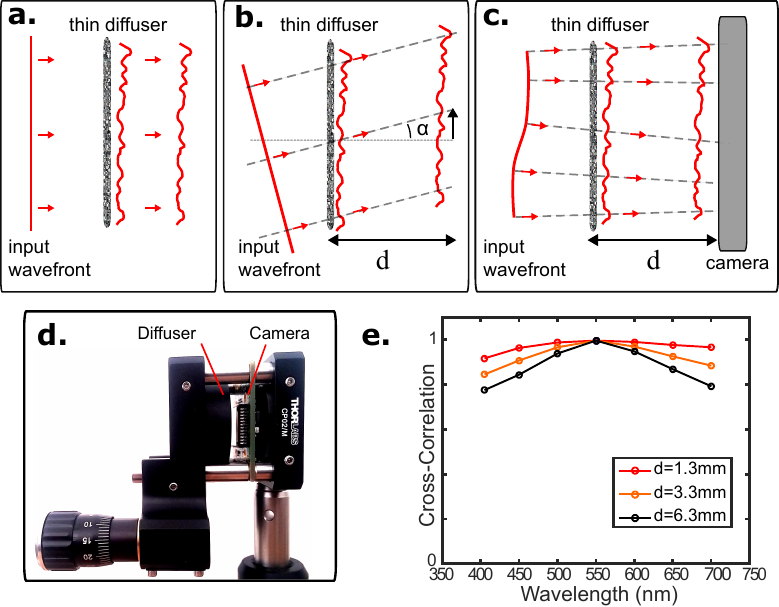}
\caption{\textbf{Principle of the thin diffuser based wavefront-sensor}. For a thin diffuser, a tip or tilt angle $\alpha$ to the impinging wavefront results in a global shift of the speckle pattern by an amount $\alpha d$, at a distance $d$. For a distorted wavefront, speckle grains are locally shifted (c). Practical implementation of the wavefront sensor (WFS) is shown in (d). This WFS exhibits broadband performances as demonstrated by the speckle correlation function as a function of wavelength for different distance $d$ (e).}
\label{fig:Fig1}
\end{figure}

\begin{figure*}[htbp]
\centering
\includegraphics[width=\linewidth]{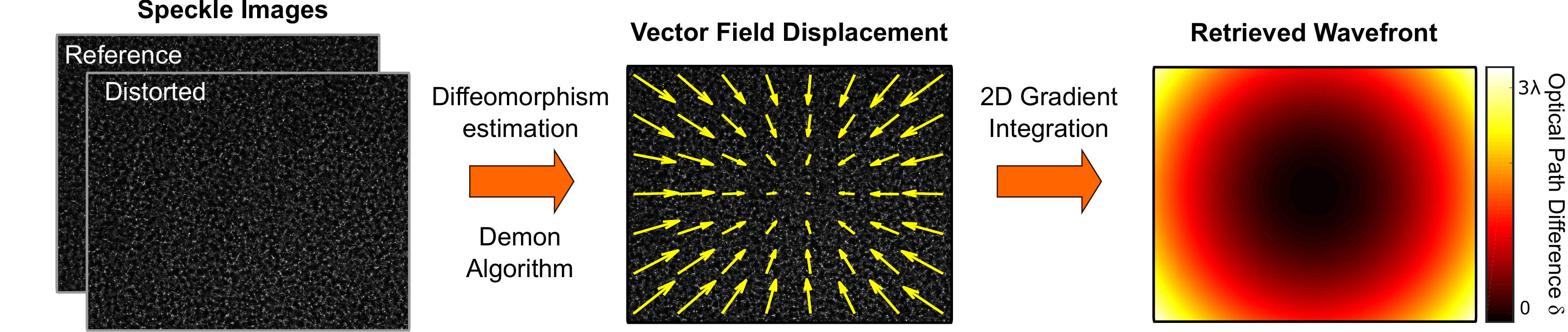}
\caption{\textbf{Wavefront reconstruction from distorted speckle maps}. The speckle pattern of the signal wavefront is compared to the speckle recorded with a flat reference thanks to a diffeomorphic algorithm. The displacement vector field is proportional to the transverse component of the local wavevector. Numerical integration then allows retrieving the phase of the field.}
\label{fig:Fig2}
\end{figure*}

Wavefront sensing is usually performed with a camera placed at the focal distance of an array of microlenses in Shack-Hartmann (SH) WFS~\cite{Shack_1971,Vdovin_OE_16} or at few millimiters of a grating in lateral shearing interferometry (LSI)~\cite{Primot_1995}. The local tip/tilt of the wavefront thus results in a shift of the spots (SH case) or fringes (LSI cases) on the camera. Comparatively to microlens arrays and gratings, diffusers have the advantage to be low-cost and can also encode the wavefront in an intensity pattern. %\textbf{Thanks to ever-increasing computer performances, the prior measurement of the transmission matrix of the diffuser by DH allows imaging~\cite{Lee2016}}. 
However the information is more difficult to retrieve because the input and output fields are connected through a complex transmission matrix whose accurate description requires a long and tedious calibration~\cite{Gigan_PRL_11,Lee2016}.
The calibration step can be avoided in the approximation of weakly disturbed beams~\cite{Bernet2011} or measuring several images to reconstruct the volumic intensity distribution of the speckle and running iterative algorithms~\cite{Almoro2006,Almoro2008}. Including sparsity constraints to the sough-for object in the error-minimization iterative algorithm -- through a Tikhonov regularization parameter -- allows retrieving the complex amplitude with a single image~\cite{Horisaki2016,Antipa2017}. %Algo Long, lourd, beaucoup a priori
In most of these techniques the diffuser is placed at a large distance from the camera. The speckle pattern thus results from the interference of a high number of waves %, ensuring efficient scrambling and sampling in the phase-space domain, but, as an interferometer, also 
making its contrast highly sensitive to the spectral bandwidth of the illumination source~\cite{Gigan_PRL_16}. Here, we point out that placing the diffuser closer to the detector reduces the number of interfering waves and thus chromatic sensitivity.
%
%Moreover, the transmission matrix of thin diffusers is almost diagonal in the real space domain: a point source at the input yields a slightly broadened point source at the output. %tu peux reformuler un peu stp -pas clair...
%Consequently, 
Moreover, the speckle generated by such a diffuser exhibits the so-called ``memory effect''~\cite{Feng1988,Freund1988}: a tip/tilt applied to the input wavefront results in a simple tip/tilt on the output wavefront (illustrated in Fig.~\ref{fig:Fig1}a and ~\ref{fig:Fig1}b). This remarkable property gave rise to speckle interferometry in astronomy \cite{LabeyrieAA70} and has recently been used to perform non-invasive imaging through diffusers~\cite{Bertolotti2012,Katz2012,Psaltis_OE_14,Psaltis_OL_16} and multi-core optical fibers~\cite{Rigneault_OL_13,Katz_OE_16}.

In this letter, we suggest the design of a new WFS by placing a thin diffuser at a short distance from a camera sensor. %The ``short'' distance is chosen so that only \textbf{a few wave interfere} to achieve broadband spectral performances. In addition, u
Using a thin diffuser ensures a large ``memory effect'' and makes that a local tip/tilt in the wavefront results in a simple shift of the speckle grains over the camera sensor, similar to SH and LSI, as illustrated in Fig.~\ref{fig:Fig1}c. The measured speckle pattern from an unknown wavefront can then be compared to a reference speckle pattern thanks to a simple and fast diffeomorphic image registration algorithm. 
The proposed WFS is shown in Fig.~\ref{fig:Fig1}d. An holographic diffuser ($\theta = 1^\circ$, Edmund Optics, York, UK) is placed at a few millimeters from a monochrome CMOS camera (DCC 1545M - 8 bits -1280x1024, Thorlabs). 
The diffuser exhibits a very large angular memory effect (correlation higher than $95\%$ for $\Delta \alpha=0.3~{\rm rad}$). 
Using the Wiener-Khintchine theorem (and relying on the Gaussian intensity distribution at infinity), we obtain the correlation width (FWHM) of the diffuser: $a=\frac{\lambda}{\theta}\frac{2\sqrt{2}\ln(2)}{\pi}$ ($a\simeq 18~ \mu{\rm m}$ at $\lambda=500~{\rm nm}$).
%This correlation width was chosen so that 
The speckle grain can then be resolved by the camera (pixel size: $5.2~\mu {\rm m}$). 
%Along the propagation axis, the correlation depth of the speckle %in the vicinity of the diffuser 
%is $d_0= 2\sqrt{2} \frac{a}{\theta} \simeq 2.8~{\rm mm}$. 
%For $d$ in the millimeter range, the speckle pattern at the camera thus results from the interference of a few waves only.
%, with phase delays scaling as $\delta \varphi_k= k  d (\theta/2)^2/2$. For a difference of wavenumber $\Delta k$, the interference patterns significantly differs for $\delta \varphi_{\Delta k}\simeq\pi$, yielding $d\geq\frac{8}{\theta^2\Delta k}$ ($\simeq 1~{\rm cm}$ for $\Delta\lambda=150~{\rm nm}$). 
%
Behind the diffuser, the speckle grain displacement vector field~${\bf s}$ on the camera is proportional to the gradient of the optical path difference (OPD) experienced by the beam:%~$\nabla_\bot \delta$: %$\left(\nabla_{x}(\textit{W}),\nabla_{y}(\textit{W})\right)$ 
 \\% ($s_{x}$, $s_{y}$) in pixel.\\ 
\begin{equation}
\nabla_\bot\ \delta=\frac{\nabla_\perp \varphi}{k_0} \simeq \frac{{\bf s}}{d}
\label{eq:gradient}
\end{equation}
where $d$ is the distance between the diffuser and the camera.
In this formula, we assumed that the local tilt angle $\alpha\simeq\tan\alpha\simeq\sin\alpha=\|\nabla_\perp \delta\|$ (illustrated in Fig.~\ref{fig:Fig1}b) is small. Furthermore the speckle is almost wavelength-independent from the shortest achievable distance $d=1.3~{\rm mm}$ up to $d=6.3~{\rm mm}$ %with correlations higher than $75\%$ over a range of $\pm 150~{\rm nm}$ 
(Fig.~\ref{fig:Fig1}e). 
The loss in speckle contrast with spectral bandwidth at large distances $d$ can be estimated in the frame of the Fresnel diffraction approximation. % (when the second order term of the phase expansion in the Fresnel-Kirschhoff's integral $\frac{\theta^4 d}{8\lambda}\ll 1$ is negligible). 
In the phase term of the diffraction integral, $d$ appears as the product $\lambda d$, resulting in an intensity profile along the propagation axis simply scaling as $1/\lambda$. Therefore, the speckle patterns for different wavelengths are identical with just an axial homothetic relation (assuming the refractive diffuser is achromatic). For a polychromatic beam, the speckle typically blurs out at distances where the axial shift ($= \frac{\Delta k}{k} d$) becomes comparable with the axial correlation depth of the speckle $d_0= 2\sqrt{2} a/\theta$ which yields: $\Delta k d=16\ln(2)/\theta^2$. %$\simeq 3.6.10^4~{\rm rad}$
For $\Delta\lambda=150~{\rm nm}$, we get a blurring distance $d\simeq 9.6~{\rm mm}$. Our WFS is thus compatible with broad-band light sources.

The wavefront reconstruction procedure is illustrated in Fig~\ref{fig:Fig2} and may be split into three steps. First, the speckle grain field displacement relatively to a reference speckle pattern recorded with a planar wavefront must be estimated. For a given distance between the diffuser and the camera, a single reference image is recorded and a trivial calibration step is performed.
Numerically, measuring the image distortion consists in a digital image correlation such as required for constraint measurements in mechanics~\cite{Pan2009}. One possibility then consists in splitting the image surface into macro-pixels containing at least one speckle grain each and to calculate the local cross-correlation between the reference image and the signal image. However, doing so has two main drawbacks: it is computationally slow and if the displacement is larger than the macro-pixel size, the shift cannot be computed. Note that such a problem is also encountered with LSI and SH \cite{Herriot2016}. Here, we propose to use the so-called ``Demon algorithm''~\cite{Thirion1998,Pennec-1999,Vercauteren2009,Lombaert2014}, optimized to perform non-rigid registration of bio-medical images. This algorithm -- implemented in the Image Processing Toolbox of Matlab (\emph{imregdemons()}) -- consists in minimizing a mean squared difference function using an iterative second order gradient descent algorithm. The displacement vector map is iteratively updated in combination with Gaussian filters of decreasing dimensions to gradually achieve a multi-scale optimization and avoid local-minima stagnation artifacts. In this regard, the combination of this multi-scale algorithm with speckle patterns optimizes the dynamic range of our WFS as compared to SH and LSI which use periodic patterns prone to spots-assignment ambiguities. We obtained the best results defining $\sigma \simeq 5~{\rm pixels}$ for Gaussian smoothing, and the algorithm converged after three iterations for three pyramid levels. Processing of the Demon algorithm on graphic card takes typical times of the order of $0.34~{\rm s}$ (NVidia Quadro M2000). % \textbf{Calculation on CPU of the displacement map only takes 0.94~s but can be speed up using GPU processing  to 0.33~s (NVidia Quadro M2000)}.

Second, once the displacement map ${\bf s}$ is known with a sub-pixel accuracy, integration of \eqref{eq:gradient} is achieved in the Fourier domain~\cite{Huang2015}. To avoid artifacts associated with periodic boundary conditions assumed by fast Fourier transform (FFT), anti-symmetrization of the gradient fields are performed before integration~\cite{Bon2012}. In addition, this methodology allows computing the tip/tilt  of the wavefront which could not be retrieved by integration otherwise.\\

\begin{figure}[htbp]
\centering
\includegraphics[width=\linewidth]{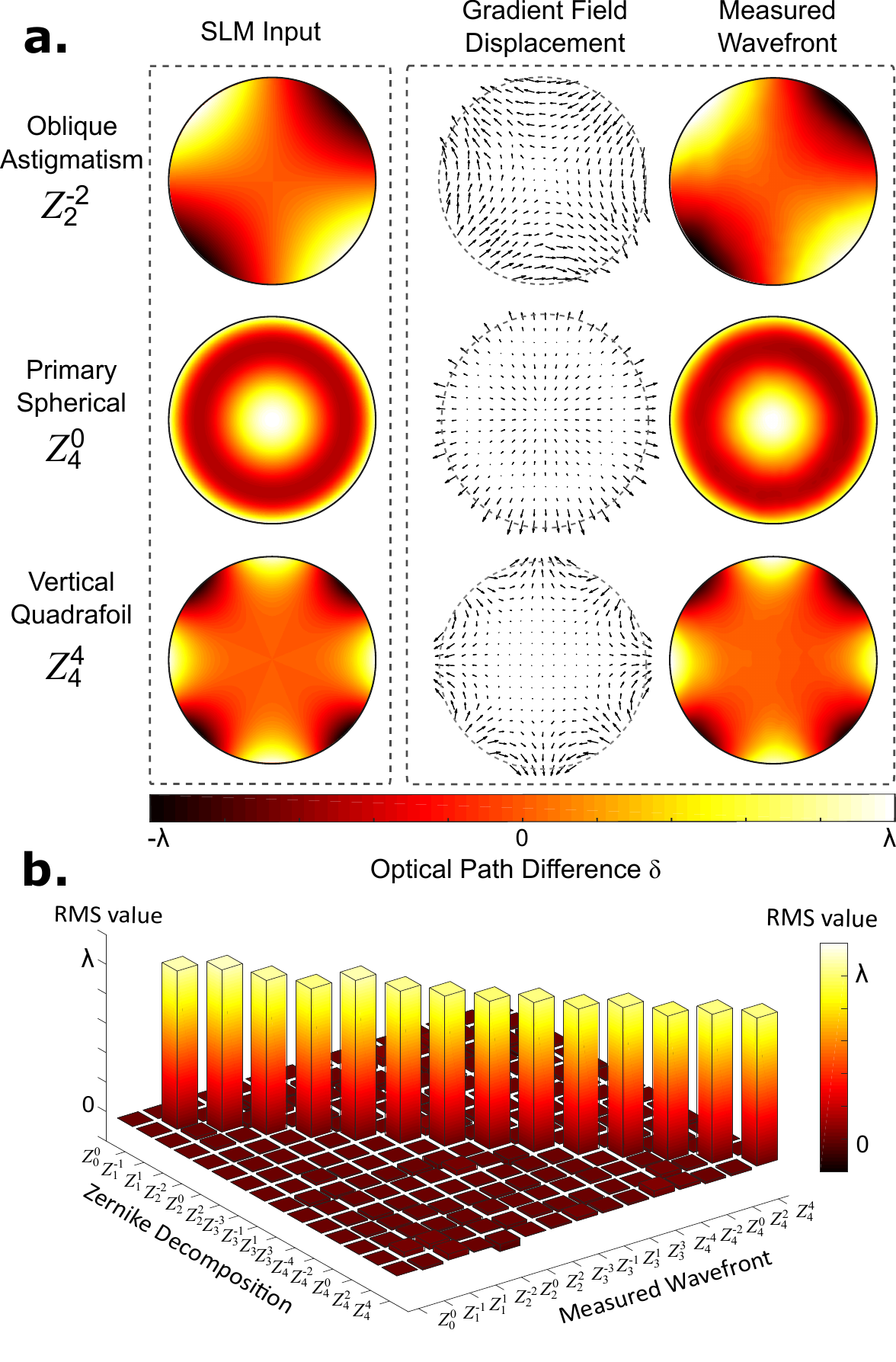}
\caption{\textbf{WFS quantitative phase measurement.} Normalized Zernike polynomials are addressed to a phase only SLM conjugated to the WFS. The measured speckle displacement map and the retrieved wavefront are shown for three Zernike polynomials \textbf{(a)}. The first fifteen Zernike modes are displayed by the SLM and the corresponding reconstructed wavefronts are decomposed on the same basis \textbf{(b)}, demonstrating quantitative phase capabilities.}
\label{fig:Fig3}
\end{figure}

Third, the coefficient of proportionality between the displacement map and the local wavevector must be precisely determined. For this calibration, a precise knowledge of the distance $d$ between the thin diffuser and the camera chip is required. This can be simply performed by imprinting a tip/tilt transform of known amplitude or placing a lens of known focal length before the system. 

To demonstrate the possibility to measure quantitatively the projection of a distorted wavefront on Zernike polynomials~\cite{Born1994}, we addressed polynomials normalized to unity to a phase-only spatial light modulator (SLM) (LCoS X10468-01, Hamamatsu, Massy, France) illuminated by a laser beam at $\lambda=532~\rm nm$, and conjugated to our WFS thanks to a telescope. A $\times 0.4$ magnification was chosen between the SLM and the camera to match sensor dimensions and the distance diffuser/camera was $d=3.11~mm$. A flat phase was first displayed on the SLM to get a reference image so that wavefront measurements are insensitive to static aberrations introduced by the optical system. %(due to the telescope, the impinging laser beam and the non-flatness of the SLM). 
Fig.~\ref{fig:Fig3}a shows the measured speckle grain displacement maps, and the corresponding reconstructed wavefronts of three Zernike polynomials. The first forteen Zernike modes addressed to the SLM (skipping the piston) were then projected in this same basis after wavefront measurement and reconstruction (Fig.~\ref{fig:Fig3}b). The orthogonality and accuracy of the reconstructed wavefronts are clearly demonstrated with a RMS error on retrieved coefficients of $\simeq 3\%$. Moreover, this value is mostly affected by the quality of the wavefront generated by the SLM and by numerical errors associated with pupil centering for modal projection. To estimate the sensitivity of the WFS, we measured the noise associated with the recording and reconstruction steps only. To do so, we recorded several times the same wavefront. Phase fluctuations of the order of $\lambda/300$ (21~mrad, or 1.8~nm at $\lambda=532~nm$) where obtained. For high sought-after sensitivities, the distance $d$ can be increased so that small tips/tilts induce speckle displacements larger than this noise level. This improved sensitivity is then obtained at the expense of the source spectral width tolerance. In the following, we demonstrate that the $d$ value also impacts the spatial resolution of the WFS.%\textbf{Dependence with d? Trade off between sensitivity and (resolution/achromaticity)?} 

\begin{figure}[htbp]
\centering
\includegraphics[width=\linewidth]{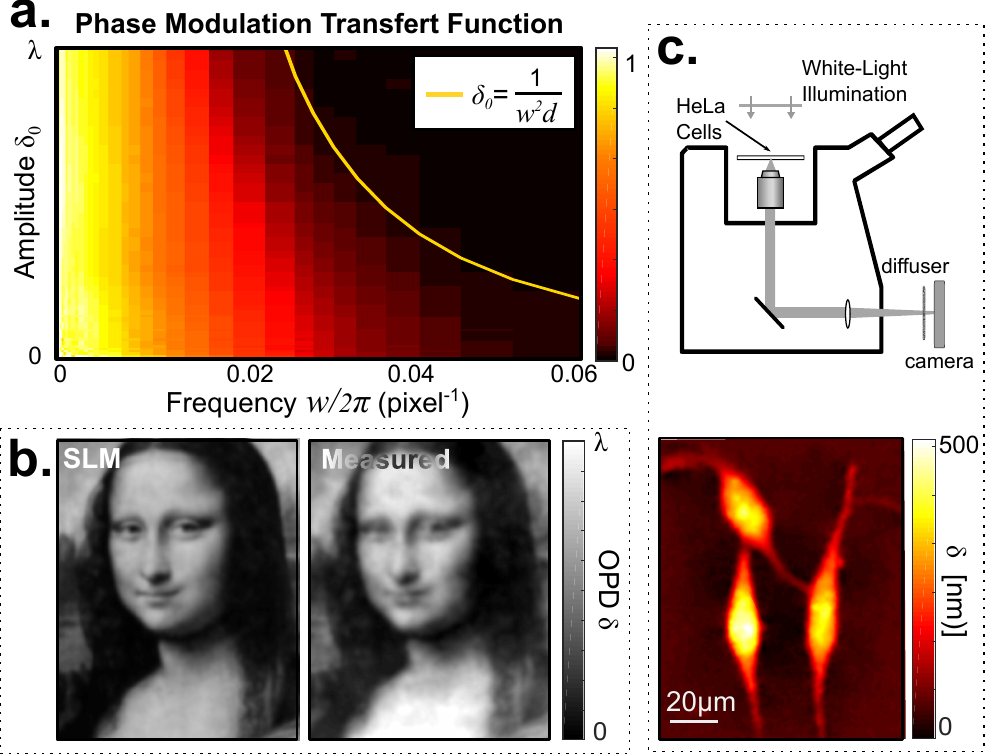}
\caption{\textbf{Quantitative phase imaging capabilities.} (a)Measured 2D phase Modulation Transfert Function of the WFS ($d=3.11$~mm). For large wavefront curvature (large amplitude, high frequency modulations region delimited by the yellow line) reconstruction fails due speckle grains overlapping.\textbf{(b)} Reconstructed phase image of \textit{Mona Lisa}. \textbf{(c)} Quantitative phase imaging of HeLa cells on a commercial microscope.}
\label{fig:Fig4}
\end{figure}

We thus now evaluate the WFS performances for imaging applications, i.e. for wavefronts exhibiting high spatial frequencies. Fig.~\ref{fig:Fig4}a shows the phase Modulation Transfert Function (MTF) of our WFS. The MTF is measured by addressing a sine-modulated OPD to the SLM: $\delta(x)=\delta_0\sin(wx)$, with various angular frequencies $w$ and amplitudes $\delta_0$. The retrieved OPD amplitude is then normalized by the input amplitude $\delta_0$. Our WFS can stand very large phase gradients thanks to the large memory effect of the thin diffuser. However, for strong curvatures of the wavefront (large coefficients in the Jacobian matrix of the gradient), speckle grains may overlap. In this case, the interference between the two overlapping speckle grains will change the speckle in a manner that cannot be described by a simple geometrical distortion. In Fig.~\ref{fig:Fig4}a, the MTF thus vanishes for simultaneously large values of $\delta_0$ and $w$. % if displaced ones towards the other of large enough amounts. 
To model this effect, let us consider the displacement $s(x)= \delta_0 w d \cos(wx)$ induced by $\delta(x)$. Speckle grains overlap at locations where $\frac{d}{dx}(x+s(x))\leq 0$, which occurs if $\delta_0 \geq 1/(w^2 d)$. %According to this model, it is then expected that image registration becomes impossible for too large values of $|\phi_0|w^2$. 
In Fig.~\ref{fig:Fig4}a, this function was plotted and qualitatively delimits the boundary of the measured MTF. This limit is also valid for other WFS like LSI and SH. %In addition, the knowledge of the MTF potentially allows optimizing the wavefront reconstruction by numerically compensating for amplitude losses at high spatial frequencies.
In the low amplitude regime ($\delta_0\ll 1$), the utmost achievable resolution is limited by the speckle grain size $a$ (here $a\simeq 18~\mu{\rm m}\simeq 3.5$~px), although the displacement map provided by the Demon Algorithm contains the same number of pixels than the camera image. In Fig.~\ref{fig:Fig4}a, the MTF limit at low $\delta_0$ values is $w/2\pi\simeq 0.05~{\rm px^{-1}}$. According to the Nyquist-Shannon criterion, this value corresponds to a maximum sampling period: $\frac{\pi}{w}\simeq 10$~pixels. To confirm this value, we also estimated the actual size of the phase-pixel by performing wavefront reconstruction by digital Image Correlation. %Splitting the image into macro-pixels and calculating the local cross-correlation between the distorted and the reference speckle image, 
We obtained reconstruction failings for macro-pixels smaller than $6\times 6$ pixels. The $1280\times 1024$ camera we use thus provides a WFS with a resolution between $13$~kpx and $36$~kpx. Fig.~\ref{fig:Fig4}b shows the reconstructed portrait of \textit{Mona Lisa} when displayed on the phase SLM, demonstrating the ability of our WFS to perform quantitative high-resolution phase imaging. 
%Reconstruction is reliable even if a slight loss in resolution can be noticed.
% Size of the gaussian smootthing 5 pixels in Demon
% Speckle grain size here? influence of d?
Finally, the WFS can be used for quantitative phase microscopy by simply placing it at the imaging port of a conventional bright-field microscope (Olympus IX71). The quasi-achromaticity of the device makes it compatible with broadband, temporally incoherent light source such as standard Halogen lamp. Closing the aperture diaphragm of the Köhler illumination as described in ref.~\citep{Bon2009} increases spatial coherence to (i) maximize the speckle contrast and (ii) allow the measurement of the OPD induced by the sample. Fig.~\ref{fig:Fig4}c shows a quantitative phase image of cultured HeLa cells. %Here, chromatic dispersion by the sample was minimized by adding a band-pass filter $580\pm 60~{\rm nm}$ in the optical path. 
  
%\textbf{Conclusion}
To conclude, we have demonstrated the possibility to make a wavefront sensor (WFS) by simply adding a thin diffuser before a camera. The proposed technique takes advantage of the so-called ``memory effect'' of the diffuser as well as the efficient diffeormorphic ``Demon'' algorithm. The proposed quantitative WFS has the advantage to be simple, broadband, compact, low-cost, high-speed and to only require trivial calibration steps. In the specific implementation we study, the sensitivity ($\lambda/300$) is limited by the 8-bits camera we used. Furthermore, the number of phase pixels ($\simeq 20$~kpx) is reduced as compared to Lateral Shearing Interferometric (LSI) techniques but it could be increased by optimizing the choice diffuser/camera and their separation distance. However, the WFS we proposed here possesses the unique property to have a tunable phase sensitivity by allowing changing the separation distance between the diffuser and the camera. A tradeoff can then be chosen by the experimentalist between sensitivity and spatial resolution. In addition, the unique signature of the speckle pattern avoids possible reconstruction artifacts associated with identification ambiguity encountered with periodic masks such as used for LSI and SH techniques. Our WFS thus offers an extensive dynamic for wavefront gradient measurement.
%Comparison with Shack Hartmann and LSI: \\
%The calibration is trivial and compared with SH or LSI, the unique signature of the speckle pattern makes aliasing artifacts absolutely impossible.
%Shack-Hartmann: Aberrant microlenses providing an unique diffraction pattern decorrelated with the others - high dynamic range! \cite{Primot_Optics_comm_2003}.%Regarde: Primot_2003 (dans WFS folder) - Description complete du shack hartmann comme un réseau! Intéressant.
%LSI:fw interference.\\
%No need of alignment, no aliasing, broad powerspectrum, Robust, relatively Fast (compatible with GPU-accelerated computing)-can be easily implemented in Matlab using the  \emph{imregdemons()}
%- non periodic pattern; high dynamic range,normal incidence illumination is not required.
%- Noise limited by 0-intensity--> 12 or 16 bits camera or log camera.\\
%Application in biology --> low-cost routine dry matter measurement

\bigskip

\noindent 
%\textbf{Funding.}  Funding information should be listed in a separate block preceding any acknowledgements. \\
\textbf{Acknowledgment.} The authors thank J{\'e}r{\^o}me Gateau for stimulating discussions.
\noindent 
% Bibliography
%\bibliography{Biblio_WFS9}

% Full bibliography added automatically for Optics Letters submissions
% Note that this extra page will not count against page length

%\ifthenelse{\equal{\journalref}{ol}}{
\clearpage
%\bibliographyfullrefs{Biblio_WFS9}  
%}{}

\end{document}